\documentclass[aps,prd,reprint,amsmath,amssymb]{revtex4-2}
\usepackage{inputenc}
\usepackage[english]{babel}
\usepackage{graphicx,color}
\usepackage{wrapfig}
\usepackage{microtype}
\usepackage{enumitem}
\usepackage{dsfont}
\usepackage{siunitx}
\usepackage{textcomp}
\usepackage{stmaryrd}
\usepackage{float}
\usepackage{hyperref}
\usepackage{titlesec}
\usepackage{comment}
\usepackage{gensymb}
\usepackage{braket}
\usepackage{mathtools} 
\usepackage{cancel}

\DeclarePairedDelimiter\autobracket{(}{)}
\newcommand{\br}[1]{\autobracket*{#1}}

\DeclarePairedDelimiter\autobrackett{[}{]}
\newcommand{\brr}[1]{\autobrackett*{#1}}

\def\va{{\boldsymbol{a}}}

\def\vr{{\boldsymbol{r}}}

\def\vp{{\boldsymbol{p}}}

\def\vr{{\boldsymbol{r}}}
\def\vt{{\boldsymbol{t}}}
\def\vv{{\boldsymbol{v}}}

\def\vR{{\boldsymbol{R}}}

\def\vA{{\boldsymbol{A}}}

\def\vB{{\boldsymbol{B}}}

\def\vR{{\boldsymbol{R}}}
\def\vT{{\boldsymbol{T}}}

\newcommand\varpm{\mathbin{\vcenter{\hbox{\oalign{\hfil$\scriptstyle+$\hfil\cr\noalign{\kern-.3ex}$\scriptscriptstyle({-})$\cr}}}}}

\hypersetup{pdfencoding=auto}

\DeclareMathOperator{\Tr}{Tr}

\begin{document}

\title{Higgs amplitude mode in ballistic  superconducting hybrid junctions}

\author{P. Vallet}
\email{pierre.vallet@u-bordeaux.fr}
\affiliation{Universit\'e de Bordeaux, Laboratoire Ondes et Mati\`ere d'Aquitaine, 351 cours de la Lib\'eration, 33405 Talence, France}

\author{J. Cayssol}
\email{jerome.cayssol@u-bordeaux.fr}
\affiliation{Universit\'e de Bordeaux, Laboratoire Ondes et Mati\`ere d'Aquitaine, 351 cours de la Lib\'eration, 33405 Talence, France}

\date{\today}

\begin{abstract}
In superconductors (SC), the Higgs amplitude mode is a coherent oscillation of the order parameter typically generated by THz laser irradiation. In this paper we propose to probe the Higgs mode using electronic transport in ballistic superconducting hybrid devices. We first confirm the existence of a non-zero amplitude mode in the clean case using the Keldysh-Eilenberger formalism. We then investigate two different device geometries, respectively a normal-insulating-superconductor (NIS) tunnel junction and a NSN junction with two transparent interfaces, the superconductor being irradiated in both situations. In the NIS case, the Higgs manifests itself in the second-order AC current response which is resonant at the Higgs frequency. In the NSN case, the DC differential conductance allows to probe the gaps dynamically generated by the Higgs mode in the Floquet spectrum. \end{abstract}

\maketitle

\section{Introduction}

Superconductivity is characterized by a spontaneous gauge symmetry breaking from the $U(1)$ group to its $\mathbb{Z}_2$ subgroup \cite{Weinberg1996Sep}. This leads to the appearance of a massive collective mode, corresponding to the coherent oscillation of the order parameter, the superconducting (SC) gap $\Delta(t)$ \cite{Anderson1963Apr, Littlewood1982Nov}. In SCs, the (Higgs) amplitude mode lies at energy $ 2 \Delta$ which corresponds to few meV, but surprisingly it was experimentally observed only in 2013 \cite{matsunaga_2013_NbN}. The reason for this late experimental evidence is that the amplitude Higgs mode is a scalar mode with no charge, and therefore no direct linear coupling to electromagnetic probes. Detecting the Higgs mode in SCs requires nonlinear coupling between light and matter only available with strong laser fields.  It is the development of THz lasers during the last decade that allowed the detection of the Higgs mode. Today, the Higgs mode has been detected in high-$T_c$ SCs in pump-probe experiments through the measurement of third harmonic generation (THG)  \cite{Higgs_dhightc,Chu2020}. Note that the presence of the Higgs mode was reported earlier through Raman spectroscopy, but in SCs showing coexistence between charge density wave order and superconductivity \cite{higgs_1980}. 

A great deal of effort has gone into understanding the role of impurities in Higgs mode excitation. It is commonly believed that, in a clean system, the Higgs mode has a negligible effect on the optical response compared to the quasiparticles (QPs) excitation (Charge Density Fluctuation) \cite{Shimano2020Mar}. Using path integral formalism, Cea \textit{et al.} \cite{Cea2016May} found that THz light cannot excite the Higgs mode due to particle-hole symmetry and that THG originates only from charge density fluctuations (CDF). However, the measurement of the THG in a NbN superconducting crystal \cite{Matsunaga2017Jul} exhibited a strongly isotropic response (as expected for the Higgs mode), contradicting the CDF hypothesis, the latter being anisotropic. To explain this experiment, several scenarios have been put forward. Phonon mediated interactions have been proposed to explain the strong response due to the Higgs mode \cite{Tsuji2016Dec}. It has been shown that impurities can drastically modify the excitation of the Higgs mode \cite{Murotani2019Jun}. Silaev, using the Eilenberger formalism, found that impurities are necessary to excite the Higgs mode with light \cite{Silaev2019Jun}. Nevertheless, Yang \textit{et al.}, using a gauge-invariant formalism, came to the opposite conclusion, namely that a finite Higgs mode could be generated even in the ideally clean case, in accordance with the Ginzburg-Landau equations \cite{Yang2018Sep, Yang2019Sep}. Vanishing of the CDF has also been demonstrated \cite{Yang2018Sep, Yang2019Sep}, in agreement with the experiment \cite{Matsunaga2017Jul}.

Most of the experimental and theoretical studies were related to the all-optical way to detect the Higgs amplitude mode in SC, typically as a THG \cite{matsunaga_2013_NbN}. Recently, a completely different route has been proposed to detect the Higgs amplitude mode, which consists in using electronic transport measurements in hybrid superconducting devices. For instance, a tunnel interface (NIS junction) between a normal metal and a dirty SC has been studied using Usadel quasiclassical equations \cite{Tang2020Jun}. In such a DC-biased NIS junction, the presence of the Higgs mode is revealed as a second harmonic in the AC current flowing through the tunnel interface. Due to progress of nanofabrication processes, NIS devices can also be build with ballistic normal parts and clean superconductors separated by interfaces ranging from tunnel to transparent ones.  

In this paper, we study clean normal-superconducting hybrid junctions. Two geometries are considered, a NIS tunnel interface and a NSN junction with highly transparent interfaces. For the DC-biased NIS junction, the signature of the Higgs mode is seen in the second harmonic in the AC current, as in the dirty case. For the NSN transparent 1D junction, the DC differential conductance provides a spectroscopy of the Floquet gaps which are dynamically induced by the Higgs amplitude mode. 

The paper is organised as follows. In Sec. \ref{section:NIS} we study the conductance of a DC-biased tunnel NIS junction when the SC amplitude mode is pumped by THz light. We first  solve the Eilenberger equations for an irradiated clean SC (\ref{section:NIS_solution}) and demonstrate that the Higgs mode can be excited even in the absence of disorder (\ref{section:NIS_Higgs}). Then we compute the second harmonic of the current flowing through the clean NIS junction  (\ref{section:NIS_current}). 
In Sec. \ref{section:NSN} we investigate the NSN ballistic junction with transparent interfaces. We solve the transport equations for this junction and obtain a DC differential conductance revealing the presence of Floquet gaps.

\section{NIS junction}\label{section:NIS}

So far the Higgs mode has been mainly studied in bulk superconductors using optical probes \cite{Shimano2020Mar}. Here we consider a NIS junction between a ballistic normal metal (N) and a clean superconductor (SC) connected by a thin insulating (I) tunnel junction. The amplitude mode of the SC is coupled to the electronic current passing through the interface and could be detected in transport experiments.

\begin{figure}[H]
    \centering
    \includegraphics[width = \columnwidth]{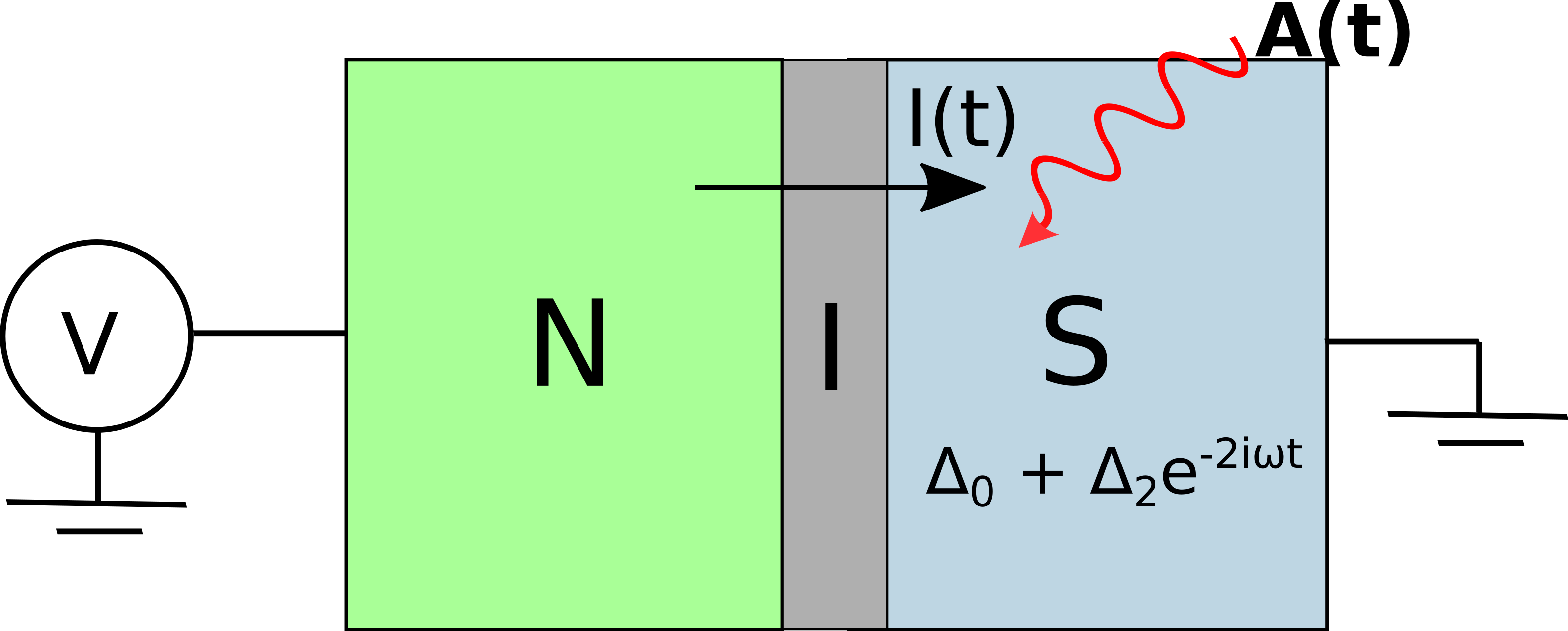}
    \caption{NIS junction. The SC region is irradiated by a monochromatic THz field $\vA(t) = \vA_0 e^{-i\omega t}$ and the normal metal is biased by a DC potential $V$ with respect to the grounded SC. Both DC and AC currents flow through the junction.}
    \label{fig:nis_schema}
\end{figure}

\subsection{Model}\label{section:NIS_model}

The superconducting region is coupled to THz light with vector potential $\vA(t) = \vA_0 e^{-i\omega t}$. The normal metal part is not irradiated but is connected to the SC by a tunnel interface. In the N region, the electrons are assumed to have a parabolic dispersion $\xi(\vp)=\vp^2/2m -\mu$, where $\mu$ is the chemical potential. The relevant momenta are close to the Fermi momentum $\vp_F$ and the dispersion can be linearized as $\xi(\vp)=  \vv_F \cdot (\vp - \vp_F)$ where $\vv_F = \vp_F/m$. A static bias potential $V$ is applied to the N region with respect to the grounded SC. Our model is closely related to the NIS junction studied in the dirty case by \cite{Tang2020Jun}, the main difference being that we treat the clean limit, for both N and SC.

 To describe the dynamics within the whole NIS structure, we use the quasi-classical (QC) limit of the Eliashberg equations \cite{Eliashberg1960Sep}, the so-called Eilenberger equations, which are valid when $\Delta / \mu \ll 1$. The Keldysh formalism with closed time contour addresses the out-of-equilibrium dynamics of the problem. We therefore introduce the Green functions in Nambu-Keldysh space \cite{Kopnin2001May}
 \begin{equation}
    \check{g} = \begin{pmatrix}
        \hat{g}^r & \hat{g}^k \\ 0 & \hat{g}^a
    \end{pmatrix} \, ,
\end{equation} 
each $\hat{g}^{i}$ being a $2\times 2$ matrix in electron/hole Nambu space and where the superscript $i=r,\ a,\ k$ stands respectively for retarded, advanced and Keldysh (or kinetic) component.
  
In the SC region, the Eilenberger equation reads \cite{Eilenberger1968,Kopnin2001May} 
\begin{equation}\label{eilenberger}
    i \left\{ \check{\tau}_3 \partial_t , \check{g} \right\} + i \brr{\Delta(t)\check{\tau} _2,\check{g}} + \brr{e\vA \cdot \vv_F \check{\tau}_3 , \check{g}} = 0 \, ,
\end{equation} 
where the Pauli matrices $\tau_i$ are embedded in the following 4x4 matrices  
\begin{equation}
    \check{\tau_i} = \begin{pmatrix}
        \tau_i & 0 \\ 0 & \tau_i 
    \end{pmatrix} \, ,
\end{equation} 
the first-order time-derivative operator acts as 
\begin{equation}
\left \{\check{\tau}_3 \partial_t,\check{g} \right\} = \check{\tau}_3 \partial_t \check{g} (t,t') + \partial_{t'} \check{g} (t,t')\check{\tau}_3 \, ,
\end{equation}
and finally the commutators have to be understood as $\left[\mathcal{O},  \check{g}\right] = \mathcal{O}(t) \check{g} (t,t')  -  \check{g} (t,t') \mathcal{O}(t')$ regarding the time arguments.

The Higgs mode is non-linearly coupled to the vector potential $\vA$. The leading nonlinear coupling is a second order one \cite{Tsuji2015Aug} with amplitude $\Delta_2$ and pulsation $2 \omega$, so that the total time-dependent order parameter reads
\begin{equation}
    \Delta(t) = \Delta_0 + \Delta_2 \, e^{-2i\omega t} \, . 
\end{equation}

In the normal metal, the Eilenberger equation reduces to 
\begin{equation}
    i \left\{ \check{\tau}_3 \partial_t , \check{g} \right\} = 0 \, ,
\end{equation} 
whose solution in Fourier space simply reads for the retarded and advanced components 
\begin{equation}
    g^r_n(\epsilon) = - g^a_n(\epsilon) = \tau_3 \, .
\end{equation}
The Keldysh component in the N region, 
\begin{align}
    g^k_n = \brr{\tanh{\br{\beta \epsilon_- /2}} - \tanh{\br{\beta \epsilon_+ /2}}} \mathds{1}\nonumber \\
    + \brr{\tanh{\br{\beta \epsilon_- /2}} + \tanh{\br{\beta \epsilon_+ /2}}} \tau_3   \, ,
\end{align} 
is related to the quasiparticle populations and contains the electrical potential $V$ via the shifted energies $\epsilon_\pm = \epsilon \pm eV$. 

The quasi-classical approximation neglects the physics at distances smaller than the superconducting coherence length, and therefore the Eilenberger equation cannot be used directly to describe the interface. Nonetheless, using microscopic Gorkov Green functions, proper boundary conditions have been established by Zaitsev for the Eilenberger Green functions \cite{Zaitsev1984}. For a tunnel junction between ballistic normal and superconducting electrodes, the Zaitsev boundary conditions  \cite{Zaitsev1984} can be expressed in the following simple form \cite{Lambert1998Feb} 
\begin{equation}\label{bc}
    \brr{\check{g}_+ - \check{g}_-}/2 = \brr{\check{g}_n,\check{g}_s} \, ,
\end{equation} where $\check{g}_\pm$ is the GF for the right (left) movers, $\check{g}_s$ (resp. $\check{g}_n$) being the GF in the SC region (resp. N region). 

The electric current can be obtain from the kinetic function \cite{Lambert1998Feb} as
\begin{equation}\label{current}
    I = \frac{G_t}{16e} \int d \epsilon \langle\Tr{\brr{\tau_3 \brr{\check{g}_n,\check{g}_s}^k}} \rangle_{\vp_F} \, ,
\end{equation} 
where $G_t$ is the tunnel conductance of the junction when the SC lead is in normal state. We denote $\langle \dots \rangle _{\vp_F} = \int d\Omega_F / 4\pi (\dots)$ the angular average over the Fermi surface. We can write the current up to the second order as the real part of 
\begin{equation}
    I(t) = I_0 + I_2 \, e^{-2i\omega t}.
\end{equation}

The second-order current can be written as a sum of two contributions $I_2 = I_V + I_H$ where $I_V$ is the current due to the second-order coupling to the vector potential only while $I_H$ is the current directly associated to the excited Higgs mode (see Appendix \ref{annex_current}).

\subsection{Second order perturbative solution}\label{section:NIS_solution}

Solving \eqref{eilenberger} for of an arbitrary shape of the time-dependent potential $\vA(t)$ is difficult. Hence, we perform a perturbative analysis with respect to the THz field amplitude, the small parameter being $A_F = e\vA_0 \cdot \vv_F $. Note that $|A_F|$ is a typical energy scale and the electromagnetic driven strength is given by the parameter $|A_F|/\omega$. Within a quasi-classical interpretation, the coupling energy $|A_F|$ corresponds to the energy gained by an electron at velocity $v_F$ in a electric field $\omega \vA_0$ during a time $1/\omega$. The GF can be expressed as a sum of functions scaling as different powers of $A_F$ as 
\begin{equation}
    \check{g}(t,t') = \check{g}_0(t,t') + \check{g}_1(t,t') + \check{g}_2(t,t') \, ,
\end{equation}
$\check{g}_i(t,t')$ being proportionnal to $A^{i}_F$. In order to solve the Eilenberger equation \eqref{eilenberger} in the $\epsilon$-space, we define the Fourier transforms

\begin{align}
    \check{g}_0(t,t') &= \int \frac{d\epsilon}{2\pi}\check{g}_0 (\epsilon) e^{-i(t'-t)\epsilon} \, ,\\
    \check{g}_1(t,t') &= \int \frac{d\epsilon}{2\pi}\check{g}_1 (\epsilon) e^{-it'\epsilon}e^{it\epsilon_1} \, ,\\
    \check{g}_2(t,t') &= \int \frac{d\epsilon}{2\pi}\check{g}_2 (\epsilon) e^{-it'\epsilon}e^{it\epsilon_2} \, ,
\end{align} with $\epsilon_n = \epsilon + n\omega$.

In the absence of irradiation, the zero-th order retarded (advanced) GF is found to be equal to \cite{Kopnin2001May} 
\begin{equation}\label{green_function0}
    g^{\alpha}_0(\epsilon) = \frac{\epsilon \tau_3 + i\Delta_0 \tau_2}{s^{\alpha}(\epsilon)},
\end{equation} with $\alpha = r,a$, where $s^{r}(\epsilon) = i \sqrt{\Delta_0^2 - \br{\epsilon + i\gamma}^2}$ and $s^{a}(\epsilon) = i \sqrt{\Delta_0^2 - \br{\epsilon - i\gamma}^2}$ with a branch-cut in the negative real line for the square root. The parameter $\gamma$ is a small positive energy necessary to impose the proper boundary condition in the $\xi$-integration. It can also be interpreted as a Dynes parameter \cite{Dynes1978Nov}, \textit{i.e.} a small phenomenological constant which describes depairing effects in the SC, induces a broadening in the optical response functions, thereby preventing an infinite resonance of the Higgs mode. 

The first-order contribution to the retarded and advanced Green function reads (see Appendix \ref{annex_eilenberger}):
\begin{equation}
\hat{g}_1^{\alpha}(\epsilon) = A_F \, \, \frac{   \tau_3 - \hat{g}_0^{\alpha}(\epsilon_1) \tau_3 \hat{g}_0^{\alpha}(\epsilon) }{s^{\alpha}(\epsilon_1) + s^{\alpha}(\epsilon)}.
\end{equation} 
The second-order contribution to the Green function is $g^\alpha_2 = g^{\alpha}_V + g^{\alpha}_H$ where
\begin{align}\label{eq:green1}
         \begin{split}
        \hat{g}_V^{\alpha}(\epsilon) &=  \frac{A_F^2}{s^\alpha_3(\epsilon)} \left[\Sigma^\alpha(\epsilon)\hat{g}^\alpha_0(\epsilon_2)\bar{\hat{g}}^\alpha_0(\epsilon_1) \hat{g}^\alpha_0(\epsilon) \right. \\ 
        &  \left. - \xi_2 - \Bar{\xi}_1 - \xi\right] \vphantom{\frac{A_F^2}{s_3(\epsilon)}},
    \end{split} \\
    \hat{g}_H^{\alpha}(\epsilon) &= \frac{i \Delta_2}{s^\alpha(\epsilon_2) + s^\alpha(\epsilon)} \brr{\tau_2 - \hat{g}^\alpha_0(\epsilon_2) \tau_2 \hat{g}^\alpha_0(\epsilon)},
\end{align}
with $\Bar{\mathcal{O}} = \tau_3 \mathcal{O}\tau_3\text{,}\ \xi_i = \epsilon_i \tau_3 {+} i\Delta_0 \tau_2$ and
\begin{align}
    s^\alpha_3(\epsilon)&=\brr{s^\alpha(\epsilon_2) + s^\alpha(\epsilon_1)}\brr{s^\alpha(\epsilon_2) + s^\alpha(\epsilon)} \brr{s^\alpha(\epsilon_1) + s^\alpha(\epsilon)}, \\
    \Sigma^\alpha(\epsilon) &= s^\alpha(\epsilon) + s^\alpha(\epsilon_1) + s^\alpha(\epsilon_2).
\end{align}
Now let us consider the Keldysh components of the Green functions which describe the non-equilibrium quasiparticle populations. In the absence of irradiation, namely at the zeroth-order in $A_F$, the stationary Keldysh GF is simply the equilibrium one \cite{Kopnin2001May}
\begin{equation}
    \hat{g}^k_0(\epsilon) = \brr{\hat{g}^r_0 - \hat{g}^a_0} \tanh{\br{\beta \epsilon/2}},
\end{equation} with $\beta = 1/k_B T$. For all orders we define $\hat{g}^k_i = \hat{g}^{\text{reg}}_i + \hat{g}^{\text{an}}_i$, where $\hat{g}_i^{\text{reg}} = g^r_i (\epsilon) \tanh{\br{\beta \epsilon/2}}- \tanh{\br{\beta \epsilon_i/2}} g^a_i (\epsilon)$.

The derivation for the other orders can be found in Appendix \ref{annex_eilenberger}. The first-order contribution reads 

\begin{align}
   \hat{g}_1^{\text{an}}(\epsilon) &= A_F\frac{\tanh{\br{\beta \epsilon_1 / 2}} - \tanh{\br{\beta \epsilon / 2}}}{s^{r}(\epsilon_1) + s^{a}(\epsilon)} \nonumber \\
   &\times \brr{\tau_3 - \hat{g}_0^r(\epsilon_1)\tau_3\hat{g}_0^a(\epsilon)} \, .
\end{align} 
The second-order Keldysh component is the sum of two terms, $\hat{g}_2^{\text{an}}(\epsilon) = \hat{g}_V^{\text{an}}(\epsilon) + \hat{g}_H^{\text{an}}(\epsilon)$, respectively given by :  

\begin{widetext}
\begin{align}\label{eq:green2}
    \begin{split}
        \hat{g}_V^{\text{an}}(\epsilon) &=  \frac{A_F^2 \brr{\tanh{\br{\beta \epsilon_2 / 2}} - \tanh{\br{\beta \epsilon_1 / 2}}}}{\brr{s^{r}(\epsilon_2) + s^{a}(\epsilon_1)}\brr{s^{r}(\epsilon_2) + s^{a}(\epsilon)} \brr{s^{a}(\epsilon_1) + s^{a}(\epsilon)}} \\
         & \times  \left[\br{s^{a}(\epsilon)+s^{a}(\epsilon_1)+s^{r}(\epsilon_2)}\hat{g}^{r}_0(\epsilon_2)\bar{\hat{g}}^{a}_0(\epsilon_1) \hat{g}^{a}_0(\epsilon)  - \xi_2 - \Bar{\xi}_1 - \xi\right] \\[2ex]
         &+ \frac{A_F^2 \brr{\tanh{\br{\beta \epsilon_1 / 2}} - \tanh{\br{\beta \epsilon / 2}}}}{\brr{s^{r}(\epsilon_2) + s^{r}(\epsilon_1)}\brr{s^{r}(\epsilon_2) + s^{a}(\epsilon)} \brr{s^{r}(\epsilon_1) + s^{a}(\epsilon)}} \\
         & \times  \left[\br{s^{a}(\epsilon)+s^{r}(\epsilon_1)+s^{r}(\epsilon_2)}\hat{g}^{r}_0(\epsilon_2)\bar{\hat{g}}^{r}_0(\epsilon_1) \hat{g}^{a}_0(\epsilon) - \xi_2 - \Bar{\xi}_1 - \xi\right] 
    \end{split}\\[2ex]
    \hat{g}_H^{\text{an}}(\epsilon) &= \frac{i \Delta_2 \brr{\tanh{\br{\beta \epsilon_2 / 2}} - \tanh{\br{\beta \epsilon / 2}}}}{s^{r}(\epsilon_2) + s^{a}(\epsilon)} \brr{\tau_2 - \hat{g}^{r}_0(\epsilon_2) \tau_2 \hat{g}^{a}_0(\epsilon)}.
\end{align}
\end{widetext}

Finally, the Higgs mode amplitude $\Delta_2$ is calculated self-consistently from the relation
\begin{equation}
    \Delta(t) = -i\frac{\lambda \pi}{4} \int_{-\omega_D}^{\omega_D} \frac{d\epsilon}{2\pi} \Tr{\brr{\langle \tau_2 g^k(\epsilon) \rangle_{\vp_F}}},
\end{equation} 
$\omega_D$ being the Debye cut-off. The equilibrium gap $\Delta_0$ depends on the temperature and we use the well-known BCS interpolation formula
\begin{equation}
    \Delta_0 (T) = \Delta_{0,0} \tanh{\brr{1.74 \sqrt{T/T_c - 1}}} \, ,
\end{equation} 
where $T_c$ is the critical temperature, $\Delta_{0,0} \equiv \Delta_0 (T=0)$.

\subsection{Higgs mode}\label{section:NIS_Higgs}

\begin{figure}
    \centering
    \includegraphics[width = 0.8\columnwidth]{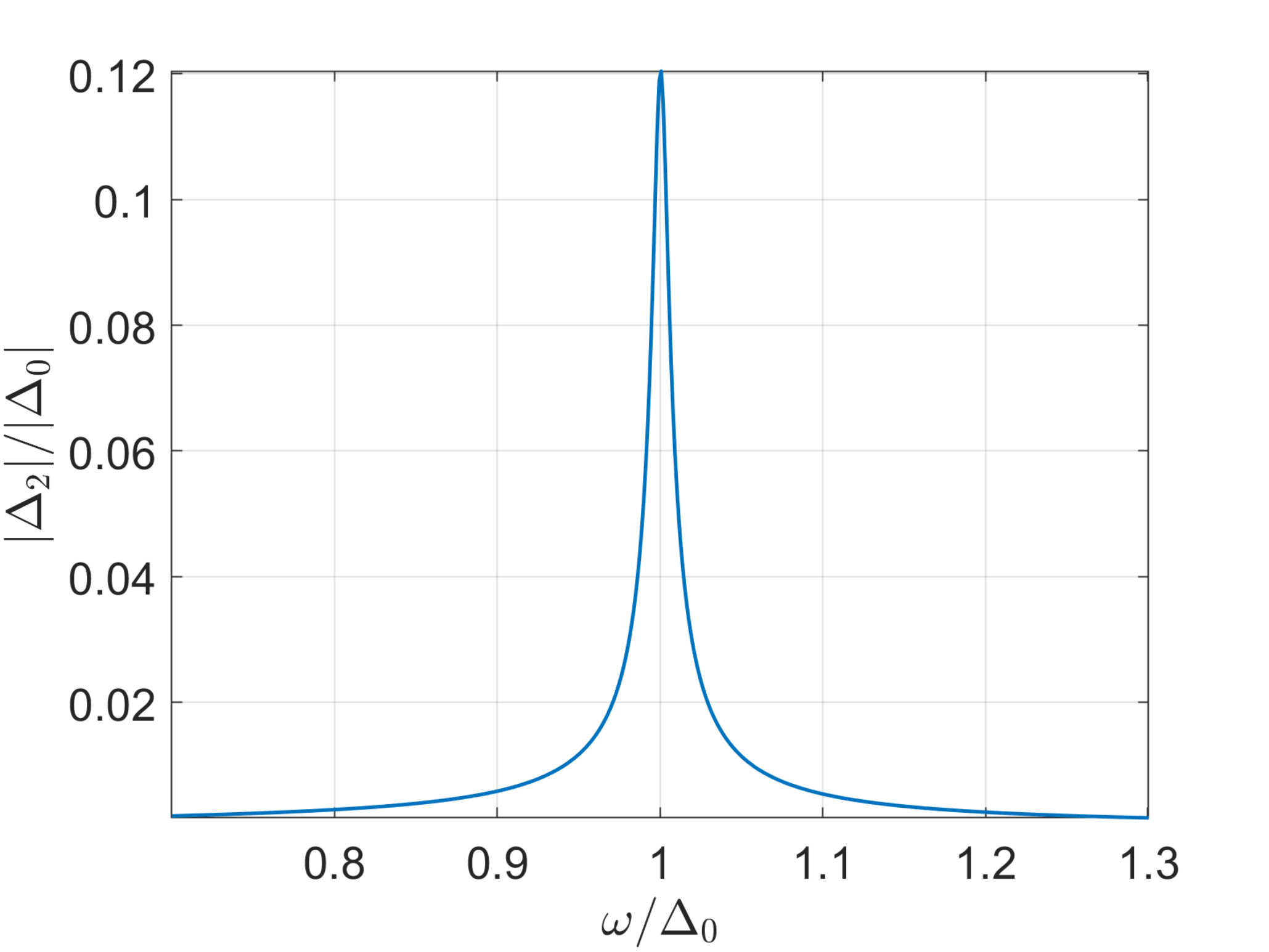}
    \caption{Higgs mode amplitude $\Delta_2 / \Delta_0$ versus the reduced light pulsation $\omega / \Delta_0$. The parameters are the amplitude $A_F^c = e |\vA_0| |\vv_F|=0.36 \, \Delta_0$, Dynes broadening $\gamma = 0.01 \, \Delta_0$, and temperature $T=0.05 \, T_c$.}
    \label{fig:higgs}
\end{figure}

A theoretical discussion is currently addressing the possibility to excite the Higgs mode in an ideally clean BCS superconductor. Using different formalisms in the clean regime, some works claimed that the Higgs mode can not be excited using optical techniques \cite{Cea2016May, Cea2018Mar,Silaev2019Jun} while others obtained a finite Higgs mode response \cite{Yang2018Sep,Yang2019Sep}. Using Keldysh real-time formalism to solve the corresponding Eilenberger equations (see previous section), we obtain a non-zero Higgs mode for all frequencies $\omega$ and observe a resonance at $\omega = \Delta$ (Fig. \ref{fig:higgs}), as expected from previous theoretical \cite{Tsuji2015Aug} results. Nevertheless, we also obtain that the Higgs mode amplitude is in principle smaller (but non zero) in clean SC than in dirty SC.  

We discuss now the differences between the clean and dirty cases, emphazising the crucial role of the anomalous contributions. First, there is a major difference between the typical energy involved in the excitations induced by irradiation : $A_F^c  = e |\vA_0| |\vv_F|$ in the clean case and $A_F^d = D \hbar \br{e\vA_0/\hbar}^2$ in the dirty case, with $D$ the Usadel diffusion constant measuring the amount of disorder. We can write the amplitude in term of a regular and anomalous function $B^\text{reg} = B^r - B^a$ and $B^\text{an}$, see Eq. \eqref{an:d2}. An interesting difference in behavior appears at this level. In the dirty case, the regular and anomalous terms share the same sign and both constructively  contribute to the Higgs mode amplitude. On the contrary, in a clean SC, those two terms have different signs and, being of same order, almost compensate each others. This sign difference explains qualitatively why the dirty case can in principle induce a stronger Higgs response from optical excitation. Still, a non-zero Higgs mode is excited in clean SC. Typically the diffusion coefficient $D\sim 1 m^2.s^{-1}$. For a Higgs mode of same amplitude in the clean and dirty case, we find that the vectors potential amplitude ratio $|\vA^d_0|/|\vA^c_0| \sim 0.1$, such that a less intense pulse in the dirty case can create a response of the same intensity as the clean case with a stronger pulse.
Note that in a recent preprint, Yang and Wu \cite{Yang2023Jan} solved partially the Eilenberger equations in the clean case. Yet, they neglected the anomalous GFs contributions to the Higgs mode (see Appendix \ref{annex_eilenberger}).

\subsection{Second-harmonic Current}\label{section:NIS_current}

\begin{figure}
  \includegraphics[width=\columnwidth]{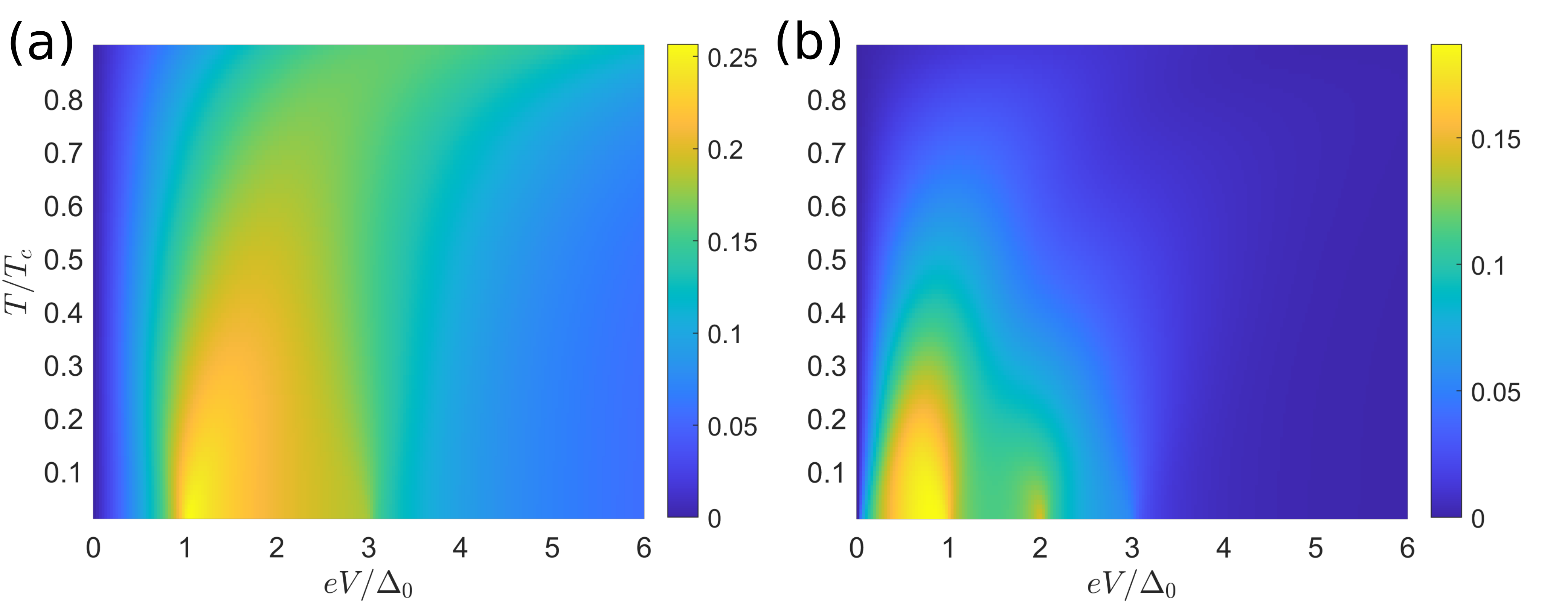}
  \caption{(a) \& (b) Maps of the ac currents $I_V$ \& $I_H$ at different reduced temperatures $T/T_c$ and bias for coupling $A^c_F = 0.36 \, \Delta_0$, $\gamma = 0.01 \, \Delta_0$. The laser pulsation is taken to be $\omega = \Delta_0(T)$. The currents are in unit of $G_t/8e$.}
  \label{fig:current_map}
\end{figure}
We now discuss the transport properties of the NIS junction and focus on the AC current at pulsation $2\omega$. The amplitude $I_2$ of this current is computed from the Green functions Eqs. \eqref{eq:green1}, \eqref{eq:green2} using Eq. (\ref{current}). We have splitted the second-order Green functions in contributions proportional to $A_F^2$ and $\Delta_2$ respectively. This results in two contributions in the current : i) a current $I_V$ directly induced by the nonlinear coupling with the electromagnetic field and ii) a Higgs current $I_H$ directly proportional to the Higgs amplitude $\Delta_2$, the formula being given in the Annex as Eq. \eqref{an:current_2}.    

Since the Higgs amplitude $\Delta_2$ is resonant at $\omega=\Delta_0(T)$, the Higgs current inherits this resonant behavior, while $I_V$ is not resonant.

\begin{figure}[H]
    \centering
  \includegraphics[width=0.8\columnwidth]{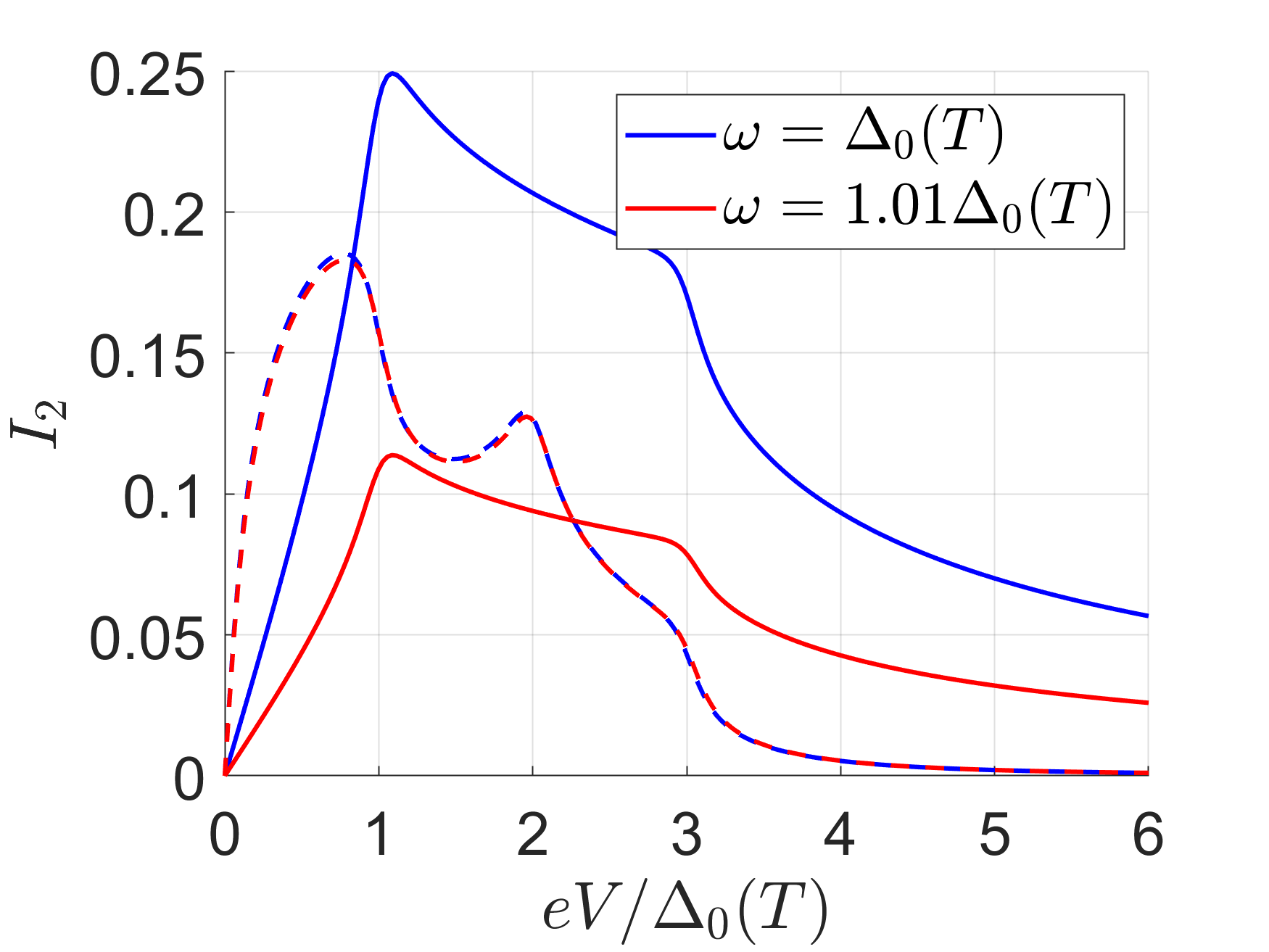}
  \caption{Amplitude of the second-harmonic currents $I_V$ (dotted line) and $I_H$ (solid line) as function of the bias potential $V$ for slightly different pulsations $\omega$ near the resonance. The Higgs current $I_H$ depends strongly on the pulsation, in contrast to $I_V$. Those amplitudes are non-monotonic for increasing DC bias voltage and decrease as $1/V$ at high bias. Here $T=0.05 \, T_c$.}
  \label{fig:current_diff_omega}
\end{figure}

Qualitatively the results are quite similar to the dirty system of \cite{Tang2020Jun}. Nonetheless interesting differences between the clean and dirty can still be seen studying the second order ac current \eqref{an:current_2}. In the dirty case, the current $I_V$ is much stronger than the Higgs current $I_H$ at the resonance. We see from Fig. \ref{fig:current_diff_omega} that in the clean case the two are of the same order, $I_H$ being still higher than $I_V$. Outside the resonance $I_V$ rapidly takes over the Higgs contribution. Qualitatively the two cases are still very similar. As in the dirty case, the current is resonant at $\omega = \Delta_0$ and it is a clear signature of the Higgs mode. More quantitatively, in the dirty case, the Higgs mode current is of the order of $\sim 30 G_t A_F^d /e$ at resonance, against $\sim 0.08 G_t A^c_F /e$ in the clean case. Knowing that for a same Higgs amplitude $A_F^d$ will be around ten times smaller than $A_F^c$ we get an estimate that, for equal Higgs amplitude mode, the current in the dirty case $I_H^d \sim 10I_H^c$ with $I_H^c$ the current in the clean system.

At resonance, the current grows quasi-linearly until the bias $eV = \Delta_0$. At this point the current starts to decrease with increasing $V$. This is due to the fact that the Higgs mode is a coherent pairing/depairing of Cooper pairs of frequency $2\Delta_0$, such that the ac current is maximum for a DC bias at the SC band edge, \textit{i.e.} for $eV=\Delta_0$  \cite{Tang2020Jun}.
As in the dirty case, $I_V$ presents different pics at frequency $\Delta_0 + n\omega$, which are signs of photon-assisted transport.

\section{NSN junction}\label{section:NSN}

In this section we study a ballistic transport problem in a clean Normal-superconducting-Normal metal (NSN) junction, where the central grounded SC part is irradiated. A signature of the Higgs mode is found in the DC differential conductance of the system.

\subsection{Model}

\begin{figure}[H]
    \centering
    \includegraphics[width = \columnwidth]{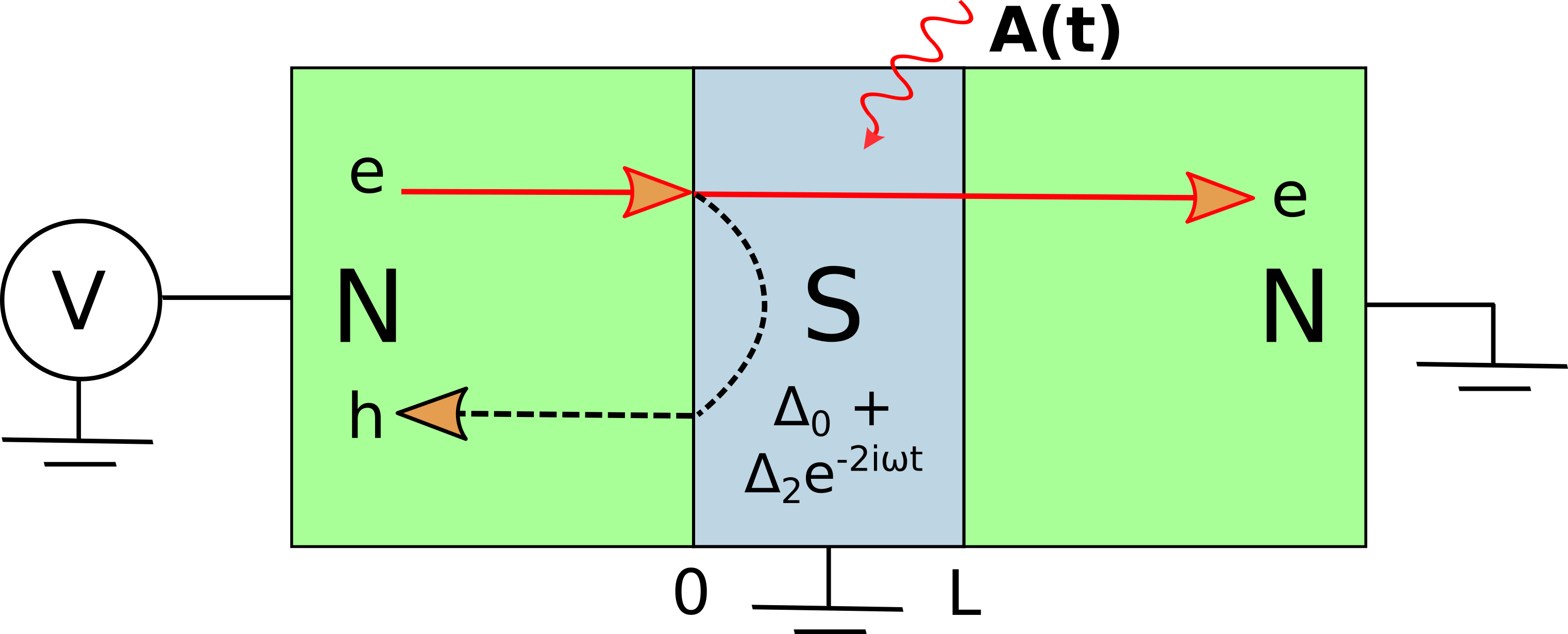}
    \caption{NSN junction. The SC is irradiated by a monochromatic THz field $\vA(t) = \vA_0 \brr{e^{i\omega t} + e^{-i\omega t}}$ and the left normal metal is biased by DC voltage $V$ with respect to SC and right N electrode. The scattering problem consists of an incident electron which can be either transmitted as an electron or Andreev-reflected as a hole.}
    \label{fig:nsn_schema}
\end{figure}

Here we propose a model consisting in a Normal-Irradiated SC-Normal metal junction. The SC has a finite length $L$. The THz light is characterized by the real vector potential $\vA(t) = \vA_0 \brr{e^{i\omega t} + e^{-i\omega t}}$. The junction is purely ballistic and we use the Bogoliubov-de Gennes (BdG) equation
\begin{equation}
    i \frac{d}{dt}\begin{pmatrix}
        u \\ v
    \end{pmatrix} = \mathcal{H}(t) \begin{pmatrix}
        u \\ v
    \end{pmatrix} 
\end{equation} 
where $u$ and $v$ are respectively the electron and hole amplitudes. The BdG Hamiltonian reads 
\begin{equation}\label{eq:ham_bdg}
    \mathcal{H} = \begin{pmatrix} H_0 - \mu & \Delta(t) \\
    \Delta ^*(t) & \mu - \mathbb{T}H_0\mathbb{T}^{-1}
    \end{pmatrix}
\end{equation} where $\Delta(t) = \Delta_0 + \Delta_2 e^{-i2\omega t}$ inside the SC and $\Delta(t)=0$ in the normal electrodes, $H_0 = \br{\hat{\vp} + e \vA(t)}^2/2m$, and $\mathbb{T}$ being the time-reversal operator. Due to the fact that $\Delta_0 / E_F \ll 1$, we can use the quasi-classical limit of this equation \cite{Kopnin2001May}

\begin{equation}\label{eq:bdg}
\begin{dcases}
    i\frac{du}{dt} = \vv_F \cdot \left( -i\partial + e \vA(t) \right) u - \mu u + \Delta(t) v \, ,\\
    i\frac{dv}{dt} = -\vv_F \cdot \left( -i\partial - e \vA(t) \right) v + \mu v + \Delta^*(t) u \, .
\end{dcases}
\end{equation}
In this approximation we completely neglect the effects of reflected electrons and crossed Andreev reflections, which is expected to be accurate in the case of fully transparent junctions \cite{Blonder1982Apr}.  

As the BdG Hamiltonian 
\begin{equation}
    \mathcal{H}(t) = \mathcal{H}(t+T), 
\end{equation} 
is periodic in time with period $T = 2\pi/\omega$, we use the Floquet formalism \cite{Floquet1883,Shirley1965May,Sambe1973Jun}. In the same way that the Bloch theorem tells that an eigenstate of a periodic in space Hamiltonian can be labeled with quasi-momentum, each state has an associated quasi-energy, \textit{i.e.}
\begin{equation}
    \Psi_\epsilon(k,t) = e^{-i\epsilon t} \Phi(k,t),
\end{equation} where $\Phi(k,t+T) = \Phi(k,t)$.

Defining the Floquet-BdG Hamiltonian as $\mathcal{H}_F(t) = \mathcal{H}(t) - i d/dt$, we find a pseudo-stationary Schrödinger equation for $\Phi(k,t)$

\begin{equation}
    \mathcal{H}_F \Phi(k,t) = \epsilon \, \Phi(k,t).
\end{equation} At this point it is usefull to introduce the following Fourier expansions  

\begin{align}
    \mathcal{H}(t) &= \sum_{n \in \mathbb{Z}} H_n e^{-in\omega t},\\
    \Phi(k,t) &= \sum_{n \in \mathbb{Z}} \Phi_n (k) e^{-in\omega t}.
\end{align}

We then have to solve an infinite number of time independent equations for the Fourier coefficients

\begin{equation}\label{eq:eqfloquet}
    \sum_{m \in \mathbb{Z}} \brr{H_{n-m} - m\omega \delta_{m,n}} \Phi _m = \varepsilon \Phi _n,
\end{equation}

with 
\begin{align}
        H_{n} &= \brr{\br{v_F k - \mu}\tau_3 + \Delta_0 \tau_1} \delta_{n,0} \nonumber \\
        &+ A_F \mathds{1} \brr{\delta_{n,1} + \delta_{n,-1}} \nonumber \\
        &+ \Delta_2\tau_+ \delta_{n,2} + \Delta_2 \tau_- \delta_{n,-2},
\end{align} where  $\tau_\pm = \tau_1 \pm i \tau_2$. In practice, to solve the equations, we choose a cut-off $N_c$ in the number of Floquet replicas. 

\subsection{Floquet spectrum}\label{sec:floquetspectrum}

\begin{figure}[H]
    \centering
    \includegraphics[width = \columnwidth]{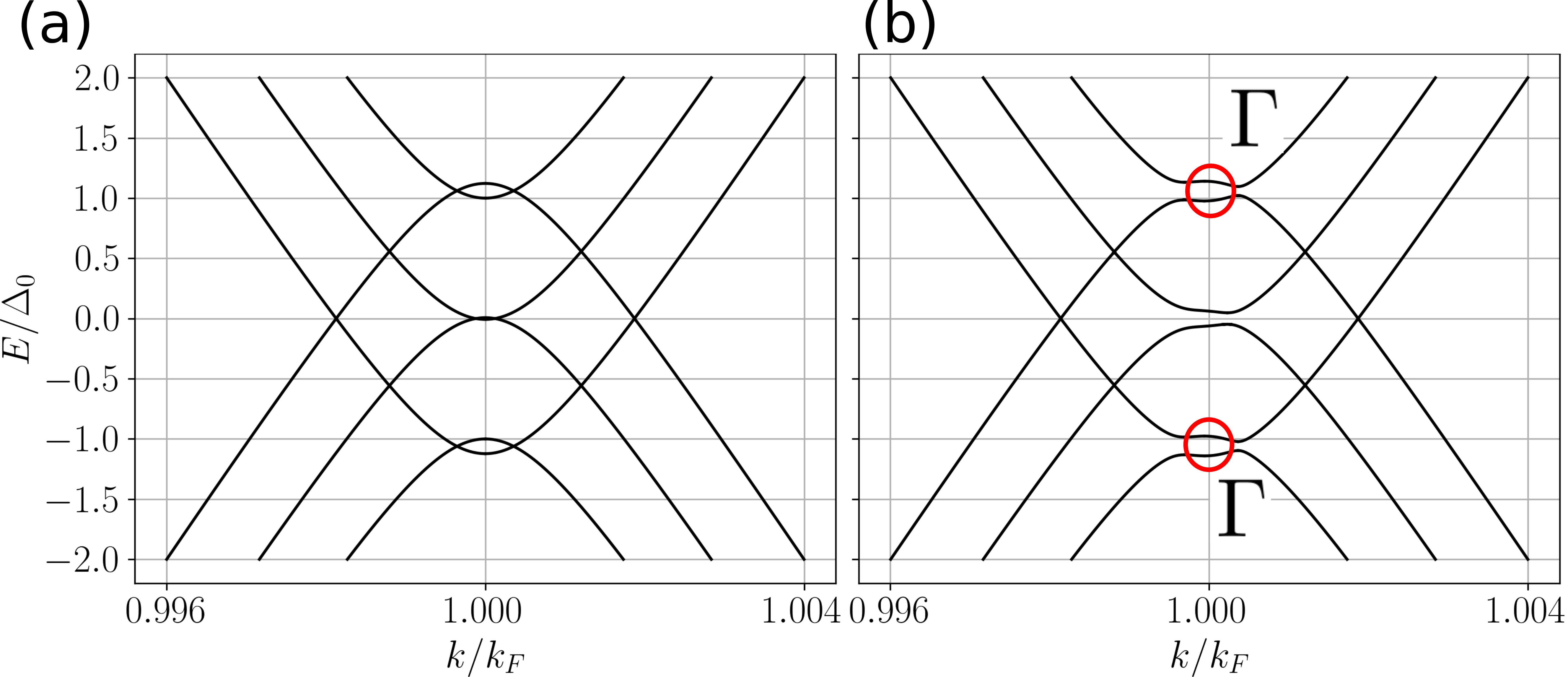}
    \caption{Floquet spectrum in the case where $A^c_F = 0.36 \Delta_{0,0}$ and \textit{(a)} $\Delta_2$ is arbitrary fixed to $0$, \textit{(b)} $\Delta_2 = 0.12 \Delta_{0,0}$ (value taken from Fig. 2). We consider the zero temperature limit $T=0$. Gaps appear only in the presence of a finite amplitude mode $\Delta_2$.}
    \label{fig:floquet_spec}
\end{figure}

The Floquet energy spectrum of the SC region presents various intercrossing bands (Fig. \ref{fig:floquet_spec}). The bands exhibits gaps at different energies. There is the superconducting gap at $\epsilon = 0$, another one at $\epsilon = \Delta_0$. This type of gap induced by THz excitation has already been discussed in the context of irradiated graphene \cite{Zhou2011Jun, Perez-Piskunow2015Apr,Atteia2017Dec}. The gaps present a rich structure depending on the electromagnetic field strength. Here we found similar results in the quasi-classical limit BdG equation. An important observation is that a gap opens only in the presence of the Higgs mode for $\epsilon = \Delta_0$. We found that the gap size $\Gamma$ at $k_F$ is to a very good approximation given by the Higgs mode amplitude $\Gamma \simeq \Delta_2$ \footnote{There are also higher order gaps at the same energy, but much smaller (\textit{e.g.} we have also another higher order gap who scales as $\propto \Delta_2^2/2$); the Higgs mode being here small compare to the other energy scales of our problem those higher order gaps will be negligible}. Of course those gaps are second order in the potential vector, as $\Delta_2 \sim A_F^{c2}$. 
In Fig. \ref{fig:floquet_spec} we compare the case with and without Higgs mode in the SC. To do that we artificially fixed $\Delta_2 = 0$ in the left figure. In this case the gaps close, showing the necessity of the Higgs mode to open a gap at $\epsilon = \Delta_0$. 

\subsection{BdG-Floquet Scattering problem }\label{sec:scattering}

We consider the following 1D scattering problem: an electron comes from the left and can be reflected as a hole or transmitted as a electron on the right. The quasi-classical approximation prevents the existence of crossed Andreev reflection in the junction or reflected electron. In the left N region ($x<0$), we have the following electron-like incident and hole-like reflected waves

\begin{align}
    \Psi_{\text{in}} &= \begin{pmatrix}
        1 \\ 0
    \end{pmatrix} \sum_n e^{ik^+_n x} e^{-i\epsilon t}e^{-in \omega t},\\
    \Psi_{\text{out}} &= \begin{pmatrix}
        0 \\ 1
    \end{pmatrix} \sum_n r_n e^{ik^-_n x} e^{-i\epsilon t}e^{-in \omega t},
\end{align} with $v_F k^\pm_n = \mu \pm (\epsilon + n \omega)$.

In the right N region ($x>L$), the transmitted wave is electron-like and reads

\begin{equation}
    \Psi_{\text{trans}} = \begin{pmatrix}
        1 \\ 0
    \end{pmatrix}  \sum_n t_n e^{ik^+_n x} e^{-i\epsilon t}e^{-in \omega t},
\end{equation} 
where $r_n$ (resp. $t_n$) is the reflection (resp. transmission) coefficient for the $n$-th Floquet level. 

Inside the SC, the solution of the Floquet-BdG equation reads 
\begin{align}
    \Psi_\text{SC} = \sum_{m} \sum_{n} a_m e^{i k_m x } \Phi^{m}_n  e^{-i \varepsilon t}e^{-i n \omega t},
\end{align} 
where $\Phi^{m}_n$ (resp. $k_m^\pm$) are the Floquet eigenvectors (resp. eigenvalues) inside the superconductor. Those excitations are coherent superpositions of electron-like and hole-like, and the spinors $\Phi^{m}_n$ are obtained by solving the following eigen-mode problem 
\begin{align}\label{eq:k}
    (v_F k-\mu) \Phi_n &= \br{\epsilon + n\omega}\tau_3 \Phi_n - i \Delta_0 \tau_2 \Phi_n \nonumber \\
        &+ A_F \tau_3 \br{\Phi_{n-1} + \Phi_{n+1}}\nonumber \\
        &- \Delta_2\tau_+ \Phi_{n-2} + \Delta_2 \tau_- \Phi_{n+2}.
\end{align}
The new matrix on the RHS of \eqref{eq:k} is not Hermitian, so nothing prevents $k$ to have some non-zero imaginary part (if this is the case, the mode is an evanescent one). The Hamiltonian is a matrix of length $2(2N_c +1) \times 2(2N_c +1)$ so the number of $k_m$ is $2(2N_c +1)$.
The full solution is obtained using the boundary condition at each interfaces, \textit{i.e.} the continuity of the spinors (see Appendix \ref{annex_floquet} for full details). This gives us the $\mathcal{S}$-matrix for this scattering problem. From the unitarity of $\mathcal{S}$ we obtain the conserved law

\begin{equation}
    \mathcal{R} + \mathcal{T} = \sum_n |r_n|^2 + |t_n|^2 = 1,
\end{equation} with $\mathcal{R}$ ($\mathcal{T}$) the total reflection (transmission) coefficient.

\subsection{DC Differential conductance}

Here we show that the differential conductance of the NSN junction is a simple way to probe and realize an electronic/transport spectroscopy of the BdG-Floquet band gaps discussed in Sec. \ref{sec:floquetspectrum} (Fig.  \ref{fig:floquet_spec}). From the $\mathcal{S}$-matrix, the DC current $I_{\text{DC}}$ is obtained using the extended Landauer-Büttiker formalism \cite{Moskalets2002Nov,Moskalets2011Sep}
\begin{align}\label{eq:current}
    I_{\text{DC}} &= \frac{e}{h} \sum_{\alpha=1}^2 \sum_n \int dE\  |\mathcal{S}_{\alpha 1}(E_n,E)|^2 \nonumber\\
    &\times \br{f_1^{\text{in}}(E-eV) - f_\alpha^{\text{out}}(E_n)}
\end{align} with $E$ the incident energy, $f_\alpha$ the distribution function in the lead $\alpha$, $V$ the bias potential. The lead $\alpha=1$ (resp. $\alpha=2$) is the normal metal on the left (resp. right) of the junction. The local differential conductance in the left part is given by \cite{Takane1992May, Anantram1996Jun, Lobos2015Jun} (see also Appendix \ref{annex_current}) 
\begin{equation}\label{eq:diff_conductance}
    G_{\text{DC}} = \frac{\partial I_{\text{DC}}}{\partial V} = \frac{e^2}{h}\brr{1 + \mathcal{R}(eV)}.
\end{equation}

\begin{figure}
    \centering
    \includegraphics[width = 0.8\columnwidth]{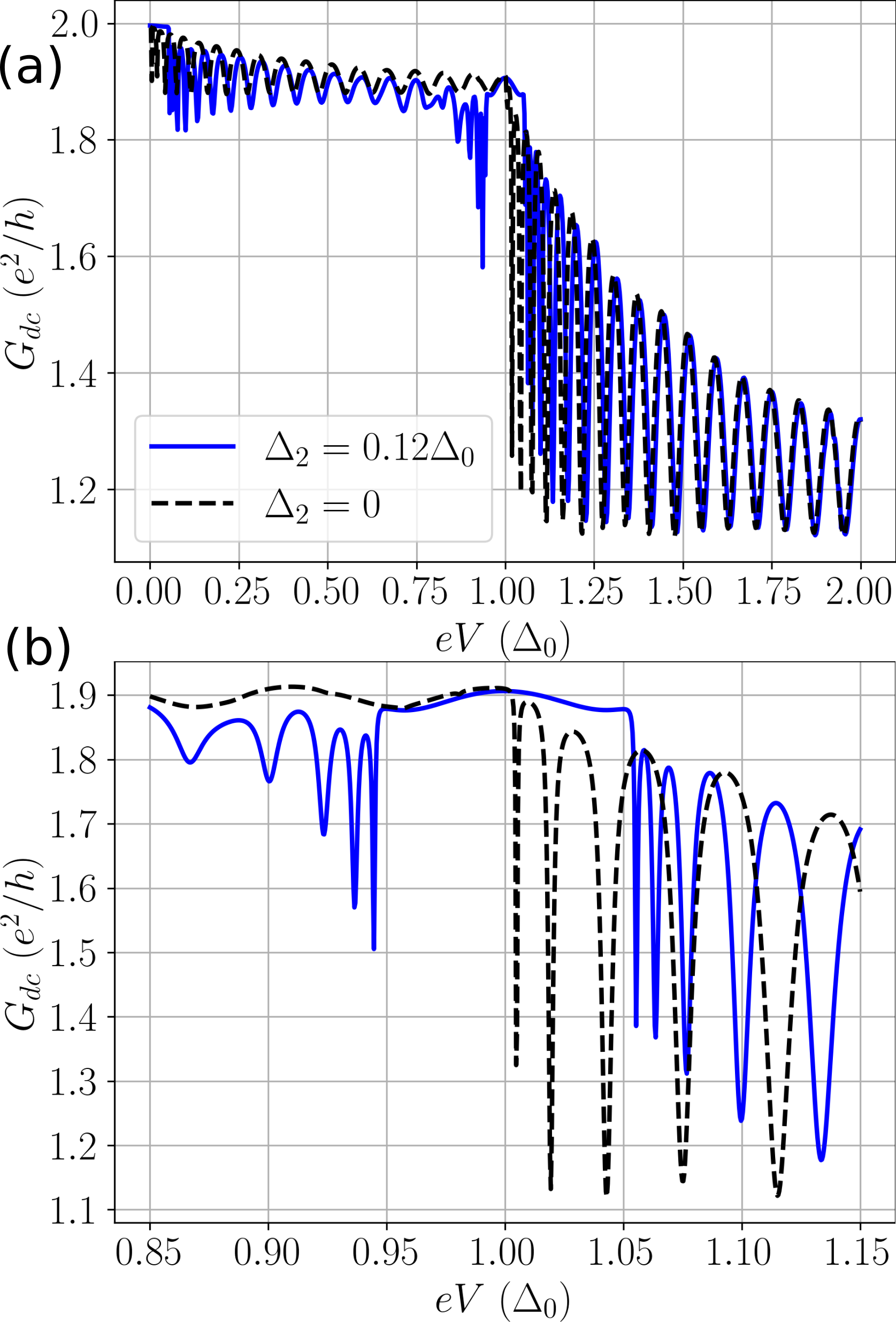}
    \caption{(a) Differential conductances as function of the DC bias $V$ with $A^c_F = 0.36 \Delta_0$, $\Delta_2 = 0.12 \Delta_0$ (blue solid line), $\Delta_2 = 0$ (black dotted line), number of Floquet replicas $N_c = 10$. (b) Zoom of the plot (a) around $eV = \Delta_0$.}
    \label{fig:conduc}
\end{figure}

 By applying a DC bias $V$ in the left normal metal, we expect a local differential conductance given by Eq. \eqref{eq:diff_conductance}. The results are given in Fig. \ref{fig:conduc}. The oscillations originate from the resonant mode inside the superconducting part. The Higgs mode appears when $eV \sim \Delta_0$. Indeed, we observe the appearance of a plateau in differential conductance centered around $\Delta_0$ and of width $\sim \Delta_2$. This plateau comes from the gap $\Gamma$. Indeed, the reflection coefficient for this range of energy will be close to $1$, as the probability for elastic transmission will be very low. For $A^c_F = 0.36 \Delta_0$, $\Delta_2 = 0.12 \Delta_0$, the first order inelastic transmission coefficient become dominant such that $|t_{-1}| \gtrsim 10 |t_0|$; elastic scattering being preferred we have $|r_0|^2 \gg |t_{-1}|^2$. Thus, $G_\text{DC} \simeq 2 e^2 /h$ in this interval and this gives the plateau. To confirm this interpretation we look at the differential conductance in the case where we artificially put $\Delta_2 = 0$. We saw before that now the gaps are closed, and we don't see any plateau anymore.

 \begin{figure}
    \centering
    \includegraphics[width = 0.9\columnwidth]{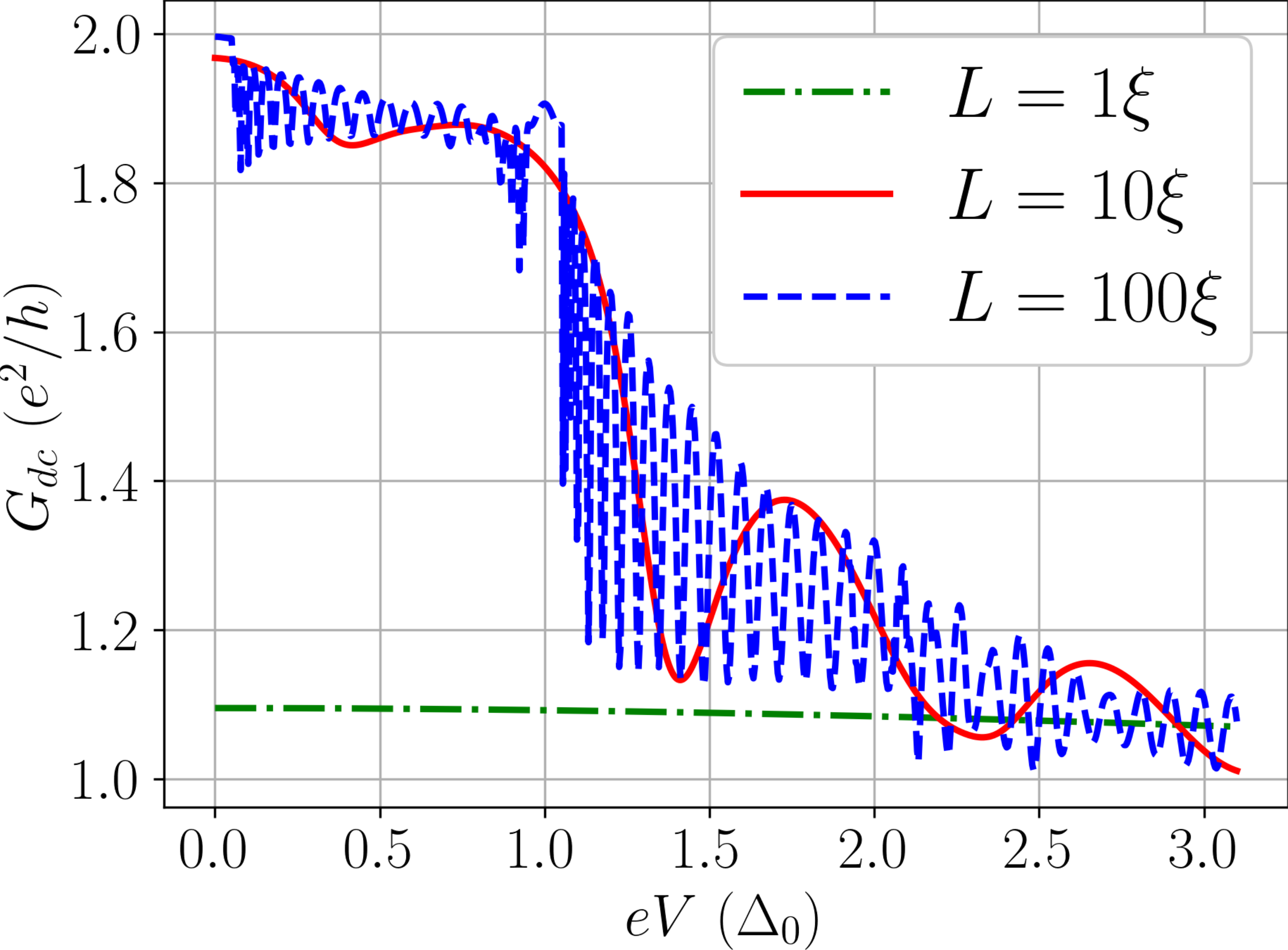}
    \caption{Differential conductances as function of the DC bias $V$ for different SC length L. The SC coherent length $\hbar v_F /\pi \Delta_0$ is noted $\xi$. The parameters are taken to be $A^c_F = 0.36 \Delta_0$, $\Delta_2 = 0.12 \Delta_0$, $N_c = 10$.}
    \label{fig:conduc_L}
\end{figure}

Looking at Fig. \ref{fig:conduc_L}, we see that the effect is only visible for large enough junction length $L$. Indeed for too small junction, the electrons injected, even inside the gap, can simply tunnel through the junction, giving this $G_\text{DC} \simeq e^2/h$ at all bias. For $\Delta_2 = 0$, The deeps of the oscillations are obtain for energy bias $eV^\text{res}_n$ that obey the resonant condition 
\begin{equation}
    eV^\text{res}_n = \sqrt{\Delta_0^2 + \br{\frac{n\pi}{L}\hbar v_F}^2}.
\end{equation} In this case the electron can tunnel through the SC and the differential conductance drops. 

The resonances observed at subgap bias are understood to come from photon assisted scattering (PAS). In this case, still for $\Delta_2 = 0$, 
\begin{equation}
    eV^\text{res}_{n,N} = \sqrt{\Delta_0^2 + \br{\frac{n\pi}{L}\hbar v_F}^2} - N\hbar \omega,
\end{equation} for $n$, $N$ such that $eV^\text{res}_{n,N} > 0$. In our case only first order PAS are visible, \textit{i.e.} the resonant bias are correctly predicted by taking $N=1$.

\subsection{Rotating wave approximation}

If the Higgs mode is present, the conductance deeps are slightly shifted compared to the case without Higgs mode, which can be explained by the modulation of the SC gap. Unfortunately, the full BdG equations \eqref{eq:bdg} are analytically intractable. Therefore, in order to get an analytical solution we consider here a simpler model where $\Delta_0 = 0$ while $\Delta_2 \neq 0$. Then the Hamiltonian \eqref{eq:ham_bdg} reduces to 
\begin{equation}
    \widetilde{\mathcal{H}}= \br{v_F p-\mu} \tau_z + v_F eA(t) \mathds{1} + \Delta_2 e^{-2i\omega t} \tau_+ +\Delta_2 e^{2i\omega t} \tau_-.
\end{equation} 
To get rid off the time dependence, we perform a "rotating frame method" by applying the unitary transformation
\begin{equation}
    \mathcal{U} = e^{it\omega \tau_z} e^{i e v_F\! \int\! A dt}
\end{equation} to the Hamiltonian. The new time-independent Hamiltonian is given by
\begin{align}
    \widetilde{\mathcal{H}}' &= \mathcal{U} \widetilde{\mathcal{H}}\mathcal{U^\dagger} + i\frac{d\mathcal{U}}{dt}\mathcal{U^\dagger} \\
    &= \br{v_F p - \mu - \omega} \tau_z + \Delta_2 \tau_x,
\end{align} whose positive eigenvalues are 
\begin{equation}\label{eq:e_rot_frame}
    \epsilon = \sqrt{\Delta_2^2 + \br{v_F p - \mu - \omega}^2}.
\end{equation} This result is formally similar to what can be found in irradiated semiconductors, and in that context the $\Delta_2$ is called a dynamical gap, induced by the electromagnetic field \cite{Galitskii1970,Syzranov2008Jul}. For this system the resonance deeps will be found at bias 
\begin{equation}
    e\widetilde{V}^{res}_n = \sqrt{\Delta_2^2 + \br{\frac{n\pi}{L}\hbar v_F - \omega}^2}.
\end{equation} This shift 
\begin{equation}\label{shiftproject}
\frac{n\pi}{L}\hbar v_F \longrightarrow \frac{n\pi}{L}\hbar v_F - \omega
\end{equation}
in the resonance bias qualitatively explains the phase shift between the resonance deeps with and without Higgs mode in Fig. \ref{fig:conduc}. We also notate that at the Higgs resonance ($\omega = \Delta_0$), a gaps appears at momentum $k \simeq k_F$ of size $2\Delta_2$. As we saw the gap reduces to $\Gamma = \Delta_2$ when we add a real $\Delta_0$ SC gap in the model. The full model with $\Delta_0 \neq 0$ is not solvable analytically. Nonetheless the general ideas are expected to still be true. 
Indeed small oscillations can be seen in Fig. \ref{fig:conduc}, around $eV \sim 0.92 \Delta_0$. Those resonances appear only in the presence of the Higgs mode and can be explained by the same sort of shifting of Eq. (\ref{shiftproject}).

\section{Conclusion}

In this work we investigated how to generate and detect the Higgs amplitude mode in ballistic superconducting hybrid devices. The Higgs mode is generated by irradiating the SC and is probed via AC or DC electronic current measurements. First, we have shown that the Higgs mode can be generated even in ideally clean SC. Then we have studied two different geometries.  We have computed the current in a NIS tunnel junction and we found a typical signature of the Higgs mode in the AC second-harmonics current, like in the dirty case. The intensity of the response is nonetheless smaller in the clean case. 
We then studied an irradiated ballistic NSN junction within Bogoliubov-de Gennes formalism. We discovered that the Higgs mode opens gaps in the Floquet band energy. Those gaps can be seen by measuring the local differential conductance of the junction around the gap energy, $eV \simeq \Delta_0 $ and their width is equal to the amplitude of the Higgs mode. The differential conductance measurements act as an electronic spectroscopy revealing the Floquet gaps dynamically generated by the presence of a finite Higgs amplitude mode.

\section{Aknowledgement}

This work was supported by the “LIGHT S\&T Graduate Program” (PIA3 Investment for the Future Program, ANR-17-EURE-0027) and GPR LIGHT.

\bibliographystyle{apsrev4-2}
\bibliography{biblio.bib}

\appendix
\section{Resolution of the Eilenberger equations}\label{annex_eilenberger}

We use the quasi-classical formalism corresponding to $\mu_F \gg \Delta_0$. From the exact Gorkov GF $\check{G}$, we define the quasi-classical one 

\begin{equation}\label{anex:quasiclassical}
    \check{g} = \frac{i}{\pi} \int d\xi_p  \, \check{G},
\end{equation}
where the integration ranges from $\xi_p =-\infty$ to $\xi_p =\infty$ and the $\check{G}$ is peaked around $\xi_p = 0$. These quasiclassical integrals typically smooth out the fast oscillating behavior of the GF at small scales.  

It fulfills the Eilenberger equation in the homogeneous case \cite{Kopnin2001May}

\begin{equation}\label{anex:eilenberger}
    i \left\{ \check{\tau}_3 \partial_t , \check{g} \right\} + i \brr{\Delta(t)\check{\tau} _2,\check{g}} + \brr{e\vA \cdot \vv_F \check{\tau}_3 , \check{g}} = 0,
\end{equation} where$\left \{\check{\tau}_3 \partial_t,\check{g} \right\} = \check{\tau}_3 \partial_t \check{g} (t,t') + \partial_{t'} \check{g} (t,t')\check{\tau}_3$, $\left[\mathcal{O},  \check{g}\right] = \mathcal{O}(t) \check{g} (t,t')  -  \check{g} (t,t') \mathcal{O}(t')$, $\vA = \vA_0 e^{-i \omega t}$, $\tau_i$ are Pauli matrices and \begin{equation}
    \check{g} = \begin{pmatrix}
        \hat{g}^r & \hat{g}^k \\ 0 & \hat{g}^a
    \end{pmatrix}
\end{equation} are the quasi-classical GFs in the $2D$ Keldysh space ; each $\hat{g}$ are $2\times 2$ matrices where respectively $r,\ a,\ k$ are for retarded, advanced and Keldysh (or kinetic) GF. Finally

\begin{equation}
    \check{\tau_i} = \begin{pmatrix}
        \tau_i & 0 \\ 0 & \tau_i 
    \end{pmatrix},
\end{equation} where $\tau_i$ are the Pauli matrices. 

The Higgs mode being a second order correction of $\Delta (t)$ we define

\begin{equation}
    \Delta (t) = \Delta_0 + \Delta_2 e^{-2i \omega t}.
\end{equation}

As usual, the gap has to be obtain self-consistently from \eqref{anex:eilenberger} ; in our case this is 

\begin{equation}\label{anex:self_consistence}
    \Delta(t) = -i\frac{\pi \lambda}{4} \Tr \brr{ \langle \tau_2 \hat{g}^k(t,t) \rangle _{\vp_F}},
\end{equation} with $\lambda$ the pairing constant and $\langle \dots \rangle _{\vp_F} = \int d\Omega_F / 4\pi (\dots)$ is the average over the Fermi surface. Eqs. \eqref{anex:eilenberger} and \eqref{anex:self_consistence} are not sufficient to get an unique solution, a normalisation condition is necessary ; it is given by 

\begin{equation}\label{anex:norm}
    \check{g} \circ \check{g}  (t,t') = \delta(t-t'),
\end{equation} with $\circ$ is the time convolution symbol. This equation cannot be solved in a simple form when $\vA$ is time-dependant. We will consider a perturbative approach : we expand the GFs in order of $\vA$ and write 

\begin{equation}
    \hat{g}(t,t') = \hat{g}_0 (t,t') + \hat{g}_1 (t,t') + \hat{g}_2 (t,t').
\end{equation} It is usefull at this point to define the Fourier transforms 

\begin{subequations}
\begin{align}
    \hat{g}_0(t,t') &= \int \frac{d\epsilon }{2\pi} e^{-i \epsilon (t-t')} \hat{g}_0 (\epsilon), \\
    \hat{g}_1(t,t') &= \int \frac{d\epsilon }{2\pi} e^{-i \epsilon_1 t} e^{i \epsilon t'} \hat{g}_1 (\epsilon), \\
    \hat{g}_2(t,t') &= \int \frac{d\epsilon }{2\pi} e^{-i \epsilon_2 t} e^{i \epsilon t'} \hat{g}_2 (\epsilon),  
\end{align}
\end{subequations} where we define $\epsilon_n = \epsilon + n \omega$. At the 0-th order the solution is

\begin{align}
    \hat{g}_0^{\alpha}(\epsilon) &= \frac{\epsilon \tau_3 + i \Delta_0 \tau_2}{s^{\alpha}(\epsilon)},\\
    s^{r(a)}(\epsilon) &= i\sqrt{\Delta_0^2 - (\epsilon \varpm i \gamma)^2},
\end{align} where the square-root branch-cut is place in the real negative line. For higher orders, we will use the following properties 
\begin{subequations}\label{anex:eq_prop}
\begin{align}
    &i \left\{ \tau_3 \partial_t , \hat{g}_i(t,t')\right\} = \int \frac{d\epsilon }{2\pi} e^{-i \epsilon_i t} e^{i \epsilon t'} \nonumber \\
    & \times \brr{\epsilon_i \tau_3 \hat{g}_i(\epsilon) - \epsilon \hat{g}_i(\epsilon) \tau_3}, \\
    &i \brr{\Delta_2 e^{-i2\omega t} \tau_2,\hat{g}_i(t,t')} = i\Delta_2 \int \frac{d\epsilon }{2\pi} e^{-i \epsilon_{i+2} t} e^{i \epsilon t'}\nonumber \\
    &\times \brr{\tau_2 \hat{g}_i(\epsilon) - \hat{g}_i(\epsilon_2) \tau_2}, \\
    &\brr{e\vA \cdot \vv_F \tau_3, \hat{g}_i(t,t')} = A_F \int \frac{d\epsilon }{2\pi} e^{-i \epsilon_{i+1} t} e^{i \epsilon t'}\nonumber \\
    &\times \brr{\tau_3 \hat{g}_i(\epsilon) - \hat{g}_i(\epsilon_1) \tau_3}, 
\end{align}    
\end{subequations} with $A_F = e \vA_0 \cdot \vv_F$. From those, it is straightforward to get the equations for the first and second order corrections of the GF 

\begin{align}
    \xi_1 \hat{g}_1(\epsilon) - \hat{g}_1(\epsilon) \xi &= A_F\brr{\hat{g}_0(\epsilon_1) \tau_3 - \tau_3 \hat{g}_0(\epsilon)},\label{anex:eq1} \\
    \xi_2 \hat{g}_2(\epsilon) - \hat{g}_2(\epsilon) \xi &= A_F \brr{\hat{g}_1(\epsilon_1) \tau_3 - \tau_3 \hat{g}_1(\epsilon)} \nonumber\\
    &+ i \Delta_2 \brr{\hat{g}_0(\epsilon_2) \tau_2 - \tau_2 \hat{g}_0(\epsilon)}. \label{anex:eq2}
\end{align}
We define the matrices $\xi_i = \epsilon_i \tau_3 + i\tau_2 \Delta_0$. To solve \eqref{anex:eq1} (\eqref{anex:eq2}), simply multiply by $\xi_1$ ($\xi_2$) from the left and $\xi$ from the right and add the two equations together. 
\begin{widetext}
    \begin{align}
\hat{g}_1^{\alpha}(\epsilon) &= \brr{\frac{A_F}{s(\epsilon_1) + s(\epsilon)} \brr{\tau_3 - \hat{g}_0(\epsilon_1) \tau_3 \hat{g}_0(\epsilon)}}^{\alpha},
    \\[2ex]
    \begin{split}
        \hat{g}_V^{\alpha}(\epsilon) &=  \left[\frac{A_F^2}{\brr{s(\epsilon_2) + s(\epsilon_1)}\brr{s(\epsilon_2) + s(\epsilon)} \brr{s(\epsilon_1) + s(\epsilon)}} \right. \\
         & \times  \left. \left[\br{s(\epsilon)+s(\epsilon_1)+s(\epsilon_2)}\hat{g}_0(\epsilon_2)\bar{\hat{g}}_0(\epsilon_1) \hat{g}_0(\epsilon) - \xi_2 - \Bar{\xi}_1 - \xi\right] \vphantom{\frac{A_F^2}{\brr{s(\epsilon_2) + s(\epsilon_1)}\brr{s(\epsilon_2) + s(\epsilon)} \brr{s(\epsilon_1) + s(\epsilon)}}} \right]^{\alpha},
    \end{split} \\
    \hat{g}_H^{\alpha}(\epsilon) &= \brr{\frac{i \Delta_2}{s(\epsilon_2) + s(\epsilon)} \brr{\tau_2 - \hat{g}_0(\epsilon_2) \tau_2 \hat{g}_0(\epsilon)}}^{\alpha}.
\end{align} 
\end{widetext} with $\Bar{\mathcal{O}} = \tau_3 \mathcal{O}\tau_3$.

To find the Keldysh function it is usefull to write the solution as a sum of regular and anomalous term 
\begin{equation}
    g^k_i (\epsilon) = g^{\text{reg}}_i(\epsilon) + g^{\text{an}}_i(\epsilon) 
\end{equation} with $g^{\text{reg}}_i(\epsilon) = g^r_i (\epsilon) h_0(\epsilon) - h_0(\epsilon_i) g^a_i (\epsilon)$ where the distribution function $h_0 (\epsilon) = \tanh{\br{\beta \epsilon / 2}}$. After some tedious but straightforward calculations we find 
\begin{align}
   \hat{g}_1^{\text{an}}(\epsilon) &= e \vA_0 \cdot \vv_F \frac{\tanh{\br{\beta \epsilon_1 / 2}} - \tanh{\br{\beta \epsilon / 2}}}{s^{r}(\epsilon_1) + s^{a}(\epsilon)} \nonumber \\
   &\times \brr{\tau_3 - \hat{g}_0^r(\epsilon_1)\tau_3\hat{g}_0^a(\epsilon)} ;
\end{align} noting $\hat{g}_2^{\text{an}}(\epsilon) = \hat{g}_V^{\text{an}}(\epsilon) + \hat{g}_H^{\text{an}}(\epsilon)$, 
\begin{widetext}
\begin{align}
    \begin{split}
        \hat{g}_V^{\text{an}}(\epsilon) &=  \frac{\br{e \vA_0 \cdot \vv_F}^2 \brr{\tanh{\br{\beta \epsilon_2 / 2}} - \tanh{\br{\beta \epsilon_1 / 2}}}}{\brr{s^{r}(\epsilon_2) + s^{a}(\epsilon_1)}\brr{s^{r}(\epsilon_2) + s^{a}(\epsilon)} \brr{s^{a}(\epsilon_1) + s^{a}(\epsilon)}} \\
         & \times  \left[\br{s^{a}(\epsilon)+s^{a}(\epsilon_1)+s^{r}(\epsilon_2)}\hat{g}^{r}_0(\epsilon_2)\bar{\hat{g}}^{a}_0(\epsilon_1) \hat{g}^{a}_0(\epsilon)  - \xi_2 - \Bar{\xi}_1 - \xi\right] \\[2ex]
         &+ \frac{\br{e \vA_0 \cdot \vv_F}^2 \brr{\tanh{\br{\beta \epsilon_1 / 2}} - \tanh{\br{\beta \epsilon / 2}}}}{\brr{s^{r}(\epsilon_2) + s^{r}(\epsilon_1)}\brr{s^{r}(\epsilon_2) + s^{a}(\epsilon)} \brr{s^{r}(\epsilon_1) + s^{a}(\epsilon)}} \\
         & \times  \left[\br{s^{a}(\epsilon)+s^{r}(\epsilon_1)+s^{r}(\epsilon_2)}\hat{g}^{r}_0(\epsilon_2)\bar{\hat{g}}^{r}_0(\epsilon_1) \hat{g}^{a}_0(\epsilon) - \xi_2 - \Bar{\xi}_1 - \xi\right] 
    \end{split}\\[2ex]
    \hat{g}_H^{\text{an}}(\epsilon) &= \frac{i \Delta_2 \brr{\tanh{\br{\beta \epsilon_2 / 2}} - \tanh{\br{\beta \epsilon / 2}}}}{s^{r}(\epsilon_2) + s^{a}(\epsilon)} \brr{\tau_2 - \hat{g}^{r}_0(\epsilon_2) \tau_2 \hat{g}^{a}_0(\epsilon)}.
\end{align}
\end{widetext}
From \eqref{anex:self_consistence} we have \begin{equation}
    \Delta_2 = \Delta_0 \frac{\int d\epsilon \Tr{\brr{ \langle \tau_2 \hat{g}^k_2(t,t) \rangle _{\vp_F}}} }{\int d\epsilon \Tr{\brr{ \langle \tau_2 \hat{g}^k_0(t,t) \rangle _{\vp_F}}} }.
\end{equation}
The previous equation can be rewritten as
\begin{equation}\label{an:d2}
    \Delta_2 = - \frac{A_F^{c2} \Delta_0 }{3}\frac{B^r - B^a + B^\text{an}}{C^r - C^a + C^\text{an}}, 
\end{equation}
with 
\begin{widetext}
\begin{align}
    B^{\alpha} &= \int d \epsilon \  b^{\alpha}(\epsilon)  \tanh{\br{\beta \epsilon_{(2)} /2}},\\
    B^\text{an} &= \int d \epsilon \   b^\text{an}_2(\epsilon) \brr{\tanh{\br{\beta \epsilon_{2} /2}} - \tanh{\br{\beta \epsilon_1/2}}} + b^\text{an}_0(\epsilon) \brr{\tanh{\br{\beta \epsilon_{1} /2}} - \tanh{\br{\beta \epsilon/2}}},\\
    C^{\alpha} &= \int d \epsilon \  c^{\alpha}(\epsilon) \tanh{\br{\beta \epsilon_{(2)} /2}} 
    - \tanh{\br{\beta \epsilon/2}} / s^{\alpha}(\epsilon), \\
    C^{\text{an}} &= \int d \epsilon \  c^{\text{an}}(\epsilon) \brr{\tanh{\br{\beta \epsilon_{2} /2}} - \tanh{\br{\beta \epsilon /2}}},
    \end{align}
where
\begin{align}
        b^{\alpha}(\epsilon) &= \left[\frac{1}{\brr{s(\epsilon_2) + s(\epsilon_1)}\brr{s(\epsilon_2) + s(\epsilon)} \brr{s(\epsilon_1) + s(\epsilon)}} \right. \nonumber \\
         & \left.\times  \left[\br{s(\epsilon)+s(\epsilon_1)+s(\epsilon_2)}\frac{\epsilon_1 \epsilon + \epsilon_2 \epsilon + \epsilon_1 \epsilon_2 + \Delta_0^2}{s(\epsilon)s(\epsilon_1)s(\epsilon_2)} -1 \right] \vphantom{\frac{1}{\brr{s(\epsilon_2) + s(\epsilon_1)}\brr{s(\epsilon_2) + s(\epsilon)} \brr{s(\epsilon_1) + s(\epsilon)}}} \right]^{\alpha},\\
         b^{\text{an}}_2 &= \frac{1}{\brr{s^{r}(\epsilon_2) + s^{a}(\epsilon_1)}\brr{s^{r}(\epsilon_2) + s^{a}(\epsilon)} \brr{s^{a}(\epsilon_1) + s^{a}(\epsilon)}} \nonumber \\
         & \times  \left[\br{s^{r}(\epsilon_2) + s^{a}(\epsilon_1) + s^{a}(\epsilon)}\frac{\epsilon_1 \epsilon + \epsilon_2 \epsilon + \epsilon_1 \epsilon_2 + \Delta_0^2}{s^{a}(\epsilon)s^{a}(\epsilon_1)s^{r}(\epsilon_2)} -1 \right],\\
         b^{\text{an}}_0 &= \frac{1}{\brr{s^{r}(\epsilon_2) + s^{r}(\epsilon_1)}\brr{s^{r}(\epsilon_2) + s^{a}(\epsilon)} \brr{s^{r}(\epsilon_1) + s^{a}(\epsilon)}} \nonumber \\
         & \times  \left[\br{s^{r}(\epsilon_2) + s^{r}(\epsilon_1) + s^{a}(\epsilon)}\frac{\epsilon_1 \epsilon + \epsilon_2 \epsilon + \epsilon_1 \epsilon_2 + \Delta_0^2}{s^{a}(\epsilon)s^{r}(\epsilon_1)s^{r}(\epsilon_2)} -1 \right],\\
         c^{\alpha}(\epsilon) &= \brr{\frac{\epsilon \epsilon_2 + \Delta_0^2 + s(\epsilon_2)s(\epsilon)}{\brr{s(\epsilon_2) + s(\epsilon)}s(\epsilon_2)s(\epsilon)}}^{\alpha},\\
         c^{\text{an}}(\epsilon) &= \frac{\epsilon \epsilon_2 + \Delta_0^2 + s^{r}(\epsilon_2)s^{a}(\epsilon)}{\brr{s^{r}(\epsilon_2) + s^{a}(\epsilon)}s^{r}(\epsilon_2)s^{a}(\epsilon)}.
\end{align}
\end{widetext}

We denote the Higgs mode as 
\begin{equation}
    \Delta_2 = \frac{F_\omega}{1 - \Pi_\omega},
\end{equation} with $F_\omega$ and $\Pi_\omega$ the amplitude and polarization functions defines in \cite{Silaev2019Jun}. From \cite{Tang2020Jun}, we see that $\Pi_\omega$ is the same in the clean and dirty case, as shown also using equilibrium Matsubara formalism in \cite{Silaev2019Jun}. The eventual presence of disorder affects only the amplitude function $F_\omega$. In our case $F_\omega \propto B^r - B^a + B^\text{an}$.

\section{Current in the NIS junction}\label{annex_current}

In a clean tunnel junction the BC can be expressed in a simple form \cite{Lambert1998Feb}

\begin{equation}
    I = \frac{G_t}{16e} \int d \epsilon \langle\Tr{\brr{\tau_3 \brr{\check{g}_n,\check{g}_s}^k}} \rangle_{\vp_F},
\end{equation} with $G_t$ the conductance of the junction. Because $\langle \check{g}_1\rangle_{\vp_F} = 0$, we can consider only the terms
\begin{align}
        \brr{\check{g}_n,\check{g}_s}_0^k &= \hat{g}^r_n \hat{g}_0^k(\epsilon) + \hat{g}^k_n(\epsilon) \hat{g}_0^a(\epsilon) \nonumber \\
        &- \hat{g}^r_0 (\epsilon) \hat{g}_n^k(\epsilon) - \hat{g}^k_0(\epsilon) \hat{g}_n^a, \\
        \brr{\check{g}_n,\check{g}_s}_2^k &= \hat{g}^r_n \hat{g}_2^k(\epsilon) + \hat{g}^k_n(\epsilon_2) \hat{g}_2^a(\epsilon) \nonumber \\
        &- \hat{g}^r_2 (\epsilon) \hat{g}_n^k(\epsilon) - \hat{g}^k_2(\epsilon) \hat{g}_n^a.
\end{align}
We now focus on the second order contribution to the current. Because $\hat{g}^k_2$ is traceless we have $\brr{\check{g}_n,\check{g}_s}_2^k = \hat{g}^k_n(\epsilon_2) \hat{g}_2^a(\epsilon) - \hat{g}^r_2 (\epsilon) \hat{g}_n^k(\epsilon).$ We now need to find the terms proportional to $\tau_3$ in $\hat{g}^{\alpha}_2$ (indeed $\tau_3 \hat{g}^k_n \propto \mathds{1} + \tau_3$). We get 
\small
\begin{align}
\begin{split}
        g^{\alpha}_{V,3}(\epsilon) &= \left[\frac{\br{e \vA_0 \cdot \vv_F}^2 }{\brr{s(\epsilon_2) + s(\epsilon_1)}\brr{s(\epsilon_2) + s(\epsilon)} \brr{s(\epsilon_1) + s(\epsilon)}} \right. \\
    &\left. \times \frac{\Sigma \epsilon \epsilon_1 \epsilon _2 + \br{\epsilon + \epsilon_1 + \epsilon_2} \brr{\Sigma \Delta_0^2 - s(\epsilon) s(\epsilon_1) s(\epsilon_2) }}{s(\epsilon) s(\epsilon_1) s(\epsilon_2)} \right]^{\alpha},
\end{split}\\
g^{\alpha}_{H,3}(\epsilon) &= \brr{\frac{2\Delta_0 \Delta_2 \br{\epsilon + \omega}}{s(\epsilon) s(\epsilon_2) \brr{s(\epsilon) + s(\epsilon_2)}}}^{\alpha},
\end{align}
\normalsize
where $\Sigma^{\alpha} = s^{\alpha}(\epsilon) + s^{\alpha}(\epsilon_1) + s^{\alpha}(\epsilon_2)$.
We can write the current $I_2 = I_V + I_H$ where 
\begin{widetext}
\begin{align}\label{an:current_2}
    I_{V(H)} &= \frac{G_t}{8e} \int d\epsilon \left[\tanh{\br{\beta (\epsilon_2 - eV) /2}} - \tanh{\br{\beta (\epsilon_2 + eV) /2}}\right] \langle \hat{g}^a_{V(H),3}\rangle_{\vp_F} \\
     &- \int d\epsilon \brr{\tanh{\br{\beta \epsilon_- /2}} - \tanh{\br{\beta \epsilon_+/2}}}  \langle \hat{g}^r_{V(H),3}\rangle_{\vp_F}. 
\end{align}
\end{widetext}

\section{S-matrix solution for the Floquet scattering in NSN junction}\label{annex_floquet}

We use the notations of Section \ref{sec:scattering} and define the vectors $\vr = \brr{r_{-N_c} \dots r_{N_c}}^T$, $\vt = \brr{t_{-N_c} \dots t_{N_c}}^T$, $\vA$ and $\vB$, the last two being the amplitudes for an incoming electron on the left and incoming hole on the right of the junction. The $\mathcal{S}$-matrix is defined by the relation 

\begin{equation}
    \begin{pmatrix}
        \vt \\ \vr
    \end{pmatrix} = \mathcal{S}\begin{pmatrix}
        \vA \\ \vB
    \end{pmatrix}.
\end{equation}

The boundary conditions, \textit{i.e.} the continuity of the spinors gives us the equations 

\begin{align}
    \begin{pmatrix}
        1\\0
    \end{pmatrix} A_n + \begin{pmatrix}
        0\\1
    \end{pmatrix} r_n &= \sum_m a_m \Phi_n^m, \\
    \begin{pmatrix}
        1\\0
    \end{pmatrix} e^{i k^+_n L}t_n + \begin{pmatrix}
        0\\1
    \end{pmatrix} e^{i k^-_n L}B_n &= \sum_m e^{i k_m L} a_m \Phi_n^m.
\end{align} From this we get 

\begin{equation}
    \begin{pmatrix}
        \vA\\ \vB e^{ik^- L}
    \end{pmatrix} = \begin{pmatrix}
        \Phi_1 \\ \Phi_2 e^{ik L}
    \end{pmatrix} \va
\end{equation} with $\br{\vB e^{ik^- L}}_n = \vB_n e^{ik^-_n L}$, $\va_n = a_n$, $(\Phi_i)_{mn}= \Phi^m_{n,i}$ and $(\Phi_i e^{ik L})_{mn}= \Phi^m_{n,i} e^{ik_m L})$  with $i$ indicating the spinor coordinate.
The $\mathcal{S}$-matrix immediately follows 

\begin{equation}
    \mathcal{S} = \begin{pmatrix}
        \Phi_1 e^{ik L} \\ \Phi_2 
    \end{pmatrix} \begin{pmatrix}
        \Phi_1 \\ \Phi_2 e^{ik L} 
    \end{pmatrix} ^{-1}.
\end{equation} We can rewrite 

\begin{equation}
    \mathcal{S} = \begin{pmatrix}
        \vR & \vR' \\ \vT & \vT'
    \end{pmatrix}
\end{equation} with $\vR$ and $\vT$ the matrices coefficients for the incoming electron from the left. For an incoming electron in Floquet band $n=0$, we will get the correct coefficients within the middle column of $\vR$ and $\vT$.

\section{Differential conductance in the NSN junction}\label{annex_g}
The current in the junction 

\begin{align}
    I_{\text{DC}} &= \frac{e}{h} \sum_{\alpha=1}^2 \sum_n \int dE\  |\mathcal{S}_{\alpha 1}(E_n,E)|^2 \nonumber\\
    &\times \br{f_1^{\text{in}}(E-eV) - f_\alpha^{\text{out}}(E_n)}
\end{align} can be written in term of the reflection and transmission coefficient

\begin{align}
    I_{\text{DC}} &= \frac{e}{h} \int dE\ \br{f_1^{\text{in}}(E-eV) - f_2^{\text{out}}(E_n)} \mathcal{T}(E) \nonumber \\ &+  \br{f_1^{\text{in}}(E-eV) - f_1^{\text{out}}(E-eV)} \mathcal{R}(E) .
\end{align} In the case of an Andreev reflection, the incident electron is reflected as a hole, such that 

\begin{equation}
    f_1^{\text{out}}(E-eV) = 1 - f_1^{\text{in}}(E-eV).
\end{equation} Only keeping the term proportional to $eV$ we get 

\begin{align}
    I_{\text{DC}} \propto \frac{e}{h} \int dE\ f_1^{\text{in}}(E-eV) \brr{\mathcal{T}(E)  + 2\mathcal{R}(E)}.
\end{align} Finally, from the conserved relation $\mathcal{R} + \mathcal{T} = 1$, we get 

\begin{align}
    G_{\text{DC}} = \frac{\partial I_{\text{DC}}}{ \partial V} = \frac{e}{h} \int dE\ \frac{\partial f_1^{\text{in}}(E-eV)}{\partial V} \brr{1+\mathcal{R}(E)}. 
\end{align} In the low temperature limit, $\partial f_1^{\text{in}}(E-eV) / \partial V = \delta(E-eV)$ and 

\begin{equation}
    G_{\text{DC}} = \frac{e^2}{h} \brr{1+\mathcal{R}(eV)}.
\end{equation}
 
\end{document}